# Quantum Tunneling of Stock Price in Range Bound Market Conditions

Ovidiu Racorean

e-mail: decontatorul@hotmail.com

**Abstract**

Applications of Quantum Tunneling effect have long gone beyond the traditional physical meaning. Initially created by Gamow to explain α-decay of nuclear particles, along the time, quantum tunneling found fertile domain of research in chemistry and recently in biology, where the new discipline of Quantum Biology emerges. The present paper extends the applicability of quantum tunneling to financial markets. In a recent paper [1] a time-independent equation for pricing the options having the underlying stock in a range bound markets is found. The equation is identical with a time-independent Schrodinger equation but incorporates elements of finance. The financial time-independent equation for option pricing is solved to explain a particular explosive violent movement of stock price in range bound markets. The aforementioned particular stock price movement is assimilated with a quantum tunneling effect. The probability of stock price to quantum tunneling out of the bounded region, known as **transmission coefficient,** is deduced. Quantum aspects of tunneling effect in financial markets are discussed. Recent evidences of price quantum tunneling in stock market are also shown.

1. **Introduction**

Quantum tunneling theory is near to celebrate the centenary anniversary. Originally, quantum tunneling was assessed in the work of Gamow [4] to explain the strange behavior of α particle in nuclear physics. Gamow ingeniously asserts that if the lower energy α particles cannot jump over the higher energy barrier of potential they must tunnel through it.

M. Born was the first to recognize that tunneling is a general result and applies to many different systems. R. Percy was the first to employ quantum tunneling for phenomenon in chemical reactions. Quantum tunneling enriches its applicability field by recent researches in biology. The success in applying tunneling to explain some DNA mutations finally emerges in creating Quantum Biology as a new and promising field of research.

The present paper emphasizes the presence of quantum tunneling effect to stock price in certain particular market conditions. Particularity of market conditions refers to stock market having prices moving in range bound, between predefined support and resistance levels that act like potential walls. Passive movements of stock price between support and resistance are sometimes evolving in explosive violent moves of high magnitude out of the range bound. These explosive price penetrations out of the range bound are explained, in the present paper by a quantum tunneling effect.

Considering pricing the options having as underlying support stocks subjected to a range bound market a time-independent equation was recently deduced [1]. The equation is identical to the one-dimensional time-independent Schrodinger equation but have financial elements incorporated. This time-independent financial equation forms the mathematical basis of price quantum tunneling effect.

The shape of the potential in the time-independent equation and the movement of the stock price within the range bound are identical with an α-particle decay process for the atomic nucleus. Following further this identity the time-independent equation is solved in the Gamow manner and the probability of stock price to tunnel out of the bounded region, known as **transmission coefficient**, is deduced. The relation for transmission coefficient contains observables from financial markets such as volatility and interest rate that probability of price tunneling is directly influenced of.

Since the transmission coefficient is very sensitive even to small changes in stock volatility, an important phenomenon is deduced - **the price tunneling effect is related with a dramatic fall in the stock volatility right prior a penetration would happen**. This effect is specific for finance and has not a counterpart in nuclear physics. Evidences of price tunneling are shown for stocks of GOOGLE and LINKEDIN, some of the most important participants in the financial market. Since the stock price quantum tunneling is very often seen in financial markets, many other examples can be easily found. A brief discussion on influences financial observables, like volatility and risk-free interest rate, has on transmission coefficient can be find in a further section of the paper.

Transmission coefficient is not just a theoretical product but an important tool that can effectively help market professionals in their investment and risk management decisions by correct timing the moment of entering the market.

Another important application in market practice, even is not the subject of the present paper, is that transmission coefficient could be employ in created trading systems based on price quantum tunneling in all financial markets that experience range bound.

2. **Explosive violent moves of stocks price in range bound markets**

Before getting involved in mathematical presentation of the quantum concepts that will be applied further, the financial background of the present paper should be briefly explained. As quantum tunneling in physics is the result of a particular state of a quantum system, so the financial tunneling of stock price occurs for a particular market conditions.

Particularity of market conditions refers to stock market being in range bound. In range bound markets price of an asset (in this case stocks) fluctuates within well that defined upper and lower boundaries known in financial markets jargon as resistance and support levels. Specification of range bound market is important since prices moves for a period of time only between support and resistance levels and replicates the behavior of a particle in a box. Support and resistance levels in market act in this case as the potential walls for the stock price.

Just like the particle could tunnel out of the box, the stock price can also break-out violently from de region bounded by the support and resistance levels. The tunneling of stock price may sound a little peculiar but is a phenomenon very familiar to traders that encounters this kind of price move almost every day.

Passive movements of stock price between support and resistance are sometimes evolving in explosive violent moves of high magnitude out of the range bound. These explosive price penetrations out of the range bound are explained, in the present paper by a quantum tunneling effect.

To better understand the concept of explosive price movement out of the range bound one suggestive example is shown in figure 1, which was taken directly from the trading software.

Figure 1

Notice the high magnitude of explosive violent price moving up after a period of passive move in range bound. These violent price movements are associated with quantum tunneling.

Bearing in mind this intuitive image of financial quantum tunneling further the attention is focus on mathematical and physical aspects of the phenomenon.

3. **Time-independent equation in options valuation**

In the recent article [1] it is recommended a time-independent equation to model the valuation of options having the underlying stock price moving in range bound. The equation in cause is identical to the one-dimensional time-independent Schrodinger equation having incorporated elements of finance and can be written as:

$$-\frac{\sigma^4}{r(\sigma^2+r)}\frac{d^2\psi_{(S)}}{dS^2} + \frac{1}{S^2}\psi_{(S)} = \frac{r}{\sigma}\psi_{(S)} \tag{1}$$

Equation (1) is deduced directly from the Black-Scholes formula and the notations are the usual in finance literature, with r being the risk-free interest rate, σ the volatility, S the price of the underlying stock and $\psi_{(S)}$ the wave function of option value.

The initial attempt of pricing options via a time-independent equation can be considered the R. Merton's work in "Theory of Rational Option Pricing" [2], work that eventually was awarded with the Nobel Prize in Economics. Merton applied newly rational option pricing concept to American put options and stated for this kind of option contracts the relation:

$$\frac{1}{2}\sigma^2 S^2 \frac{d^2w}{dS^s} + rS\frac{dw}{dS} - rw = 0, \qquad (2)$$

where w is the theoretical price of the stock option. The relation (2) is directly derived from the Black-Scholes equation:

$$\frac{\partial w}{\partial t} = -\frac{1}{2}\sigma^2 S^2 \frac{\partial^2 w}{\partial S^s} - rS\frac{\partial w}{\partial S} + rw, \qquad (3)$$

by taking the boundary condition $t \to \infty$ in which case $\frac{\partial w}{\partial t} \to 0$.

From this point on Merton's work concentrates in finding solutions of the equation for this time-independent equation, work that is not the subject of the present paper. The accent of the present paper is put on the shape of equation (2) which by the simple substitution $\ln w = \ln \psi_{(S)} - \frac{1}{2}\int \frac{2r}{\sigma^2 S} dS$ and rearranging the terms becomes:

$$-\frac{\sigma^4}{r(\sigma^2+r)} \frac{d^2\psi_{(S)}}{dS^2} + \frac{1}{S^2}\psi_{(S)} = 0. \qquad (4)$$

which clearly is a time-independent Schrodinger type equation for a particle of zero energy subject to the potential

$$V_{(S)} = \frac{1}{S^2}. \qquad (5)$$

It can be easily seen that equation (4) is a simplified form of equation (1) in the limit of a vanishing right hand term.

The shape of the potential $V_{(S)}$ is of paramount importance for the evolution of $\psi_{(S)}$ function. To evaluate the potential $V_{(S)}$ first assume that option strike price is fixed on the resistance level and the support level is set at the zeroth of the graph as is shown in Figure 2, bellow:

Figure 2

The strike price is noted K and in this particular case it equals the width of the region between support and resistance.

Notice that the potential line never cross the S axe, it is tangential to it at infinity. This image suggest that a stock price trapped in a range bound market between support and resistance levels will never be able to penetrate out of this region.

The market practice demonstrates that although the market could remain an extended period of time in range bound, price will always penetrate out. To take into account the empirical evidences on the market the equation (4) must be modified in the way that the right hand term must not vanish and should equal a constant noted **λ**:

$$-\frac{\sigma^4}{r(\sigma^2+r)}\frac{d^2\psi_{(S)}}{dS^2} + \frac{1}{S^2}\psi_{(S)} = \lambda\psi_{(S)} \qquad (6)$$

The intuitive differences between equation (4) and equation (6) can be depicted in Figure 3 that shows the potential along with different values for **λ**:

Figure 3

Notice that for $\lambda_3$, where $\lambda > V_0$ and for the Merton's hypothesis of a vanishing **λ**, **λ** is not intersecting the potential line. In these cases market is trending or if the stock price is trapped between support and resistance will remain indefinitely in this bounded region. In either situation the further discussion did not apply.

Notice also, the intersection of $\lambda_1$ and $\lambda_2$ with the walls of the potential in the condition $\lambda < V_0$. This time the stock price is moving between support and resistance and has a nonzero probability of quantum tunneling.

Setting the value of the right had term is mainly intuitively and is based on assessments of market and macroeconomic factors affecting option price in time. The extended discussion in finding **λ** values can be found in [1]. Here I will briefly sketch the main results to have the image of the process prior to explain the tunneling effect.

Using the method of separation of variables, solving equation (3) reduces to find the solution $w_{(S,t)}$ that is a product of two functions with only one variable:

$$w_{(S,t)} = \phi_{(S)}\varphi_{(t)} \qquad (7)$$

in this case time and stock price.

Putting this into Black-Scholes equation, dropping the partial derivative in favor of ordinary derivative and rearranging the terms yields:

$$\frac{d\varphi_{(t)}}{dt}\frac{1}{\varphi_{(t)}} = -\frac{1}{2}\sigma^2 S^2 \frac{d^2\phi_{(S)}}{dS^2}\frac{1}{\phi_{(S)}} - rS\frac{d\phi_{(S)}}{dS}\frac{1}{\phi_{(S)}} + r \qquad (8)$$

It follows from this, that both sides of equation (3) must equal a constant:

$$\frac{d\varphi_{(t)}}{dt}\frac{1}{\varphi_{(t)}} = \lambda \qquad (9)$$

$$-\frac{1}{2}\sigma^2 S^2 \frac{d^2\phi_{(S)}}{dS^2}\frac{1}{\phi_{(S)}} - rS\frac{d\phi_{(S)}}{dS}\frac{1}{\phi_{(S)}} + r = \lambda \qquad (10)$$

with **λ** constant.

From the time-dependent equation (9) it is stated in [1] that **λ** constant must have the form:

$$\lambda = \frac{r}{\sigma} \qquad (11)$$

Inserting this result in the time-independent equation (10), the equation (1) is recovered.

This financial time-independent equation form the basis of stock price analysis exactly in the same manner time-independent Schrodinger equation serves the quantum particle physics.

## 4. Probability of price tunneling the price potential barrier

Solutions of financial time-independent equation (1) must be analyzed taken into account the context of options price evolution. From this point of view the behavior of the $\psi_{(S)}$ function could be studied in three separate regions – price of the option are situated in the range bound, price tunneling through the potential wall, and price outside the bounded region.

Figure 4 shows the trajectory of a stock price tunneling the wall bounded by resistance level and the behavior of the $\psi_{(S)}$ function; the strike price is also situated.

Figure 4

The time-independent formula will be further solved and discussed separately for every of this regions, as all physics graduates knows from courses. The mathematics of deduction the transmission coefficient can be found in every quantum physics course and is presented here just to see where and how financial quantities fit in.

To the left of the potential wall, in the Region I, function is oscillating, price is moving between support and resistance levels, with the amplitude A:

$$\psi_1 = A e^{ikS} + B e^{-ikS} \qquad (12)$$

with $k = \sqrt{\frac{r}{\sigma^4}(\sigma^2 + r)\lambda}$

Inside the barrier wall, in the region II, function is exponential:

$$\psi_2 = C e^{qS} + D e^{-qS} \qquad (13)$$

with $q = \sqrt{\frac{r}{\sigma^4}(\sigma^2 + r)(V_{(S)} - \lambda)}$

In this region $V_{(S)} - \lambda$ is negative so that $\psi_{(S)}$ function is decreasing exponentially. In the equation (23) only the negative exponential term is retain:

$$\psi_2 = D e^{-qS} \qquad (14)$$

To the right of the barrier, in the region III, the function is oscillating again, this time with a smaller amplitude F:

$$\psi_3 = Fe^{ikS} \quad (15)$$

The probability that the price is to the left of the wall is proportional to $A^2$ and the probability that the price is to the right of the wall is proportional to $F^2$. Therefore, there is a nonzero probability T that the price striking the potential wall from the left will escape to the right:

$$T = \frac{|F|^2}{|A|^2} \quad (16)$$

It can be seen from figure 2 that F is smaller than A due to the exponential decrease of $\psi_{(S)}$ function and with a good approximation it would be find that:

$$\frac{|F|}{|A|} \approx \frac{e^{-qS_1}}{e^{-qK}} = e^{-qd}, \quad (17)$$

with d being the thickness of the potential wall.

The probability of price tunneling through the potential wall, also known as transmission coefficient, is:

$$T = e^{-2qd} \quad (18)$$

Notice the dependency of transmission coefficient to wall thickness, which make it extremely sensitive to changes in d value. As d increases the probability of tunneling become accordingly smaller and vice-versa as d decreases the T increase.

5. **Particularities of financial quantum tunneling**

To closely relate the probability of tunneling to financial quantities the width of potential wall must be first deduced.

Once the stock price penetrates the wall, the magnitude of the move can be anticipated knowing the **λ** constant value. Figure 3 shows that at the boundary of the wall **λ** must equal the potential $V_r$, which means that, at this point, the stock price will be:

$$S_r = \sqrt{\frac{1}{\lambda}} \quad (19)$$

with r = 1,2.

The penetration distance is then $d = S_r - K$. In other words, the price of underlying stock will tunnel through the wall until it reaches the $S_r$ price, penetrating through a distance:

$$d = \sqrt{\frac{\sigma}{r}} - K. \quad (20)$$

To illustrate this important result, take the strike price $K$ of a certain stock at 2,40 USD, interest rate of 3% and volatility $\sigma$ at say 47%, so that the $\lambda$ will be 0,064. From eq. 20 is trivial to derive the stock price $S_r$ at the end of the tunnel as being 3,95 USD. The price will penetrate the distance 1,55.

It can be seen in Figure 3 that the higher $\lambda$ is, the tinnier the potential wall width and the lowest $\lambda$ is, the bigger width of the wall will be.

Accounting the two parameters $\lambda$ consists of, risk-free interest rate r and volatility, it can be said that maintaining the volatility constant:

- The higher the interest rate is, the tinnier the width of potential wall that stock price should tunnel through. Accordingly the tunneling probability has high values.
- If the interest rate is low, the width of the wall grows and to penetrate the wall stock price should make big moves.

A simple example of how values of risk-free interest rate affect the distance of penetration and implicitly the probability of price tunneling is analyzed. Maintaining the same value of the strike price as in the latter examples, $K = 2,40$ USD and considering a constant volatility, varying the values of the interest rate, different values of transmission coefficient and penetrating distances are shown in the Table1:

Table1

Notice that keeping volatility constant, as the interest rate increase, the probability of tunneling is increasing due to the decreasing of wall width the price should penetrate.

Financial literature retains many different concepts for stock volatility but two are the most usually modalities of computing the fluctuation of stock prices- historical volatility and implied volatility.

Historical volatility also known as statistical volatility is the standard deviation of stock price movements. The result is annualized and remains mainly constant. From this perspective there are stocks that are more volatile and consequently have a smaller probability of tunneling when the price is trapped in range bound. On the other hand, there are stocks with low historical volatility. The high volatility stocks are more probable to overpass the exercised price, but as a consequence of equation (21), are less involved in quantum tunneling.

Considering the same example as latter with the exercised price $K = 2,40$ USD, this time maintaining the interest rate constant and varying the volatility values, different values of transmission coefficient and penetrating distances are shown in the Table 2.

Table 2

It can be notice that keeping interest rate constant, as the volatility raise, the transmission coefficient diminishes due to higher width of the wall the price should penetrate.

Implied volatility, the second modality of assessing price fluctuations, adds another important effect directly correlated with price tunneling that will be explored further. The particularity in this case reside in that implied volatility is computed any time a change in the price of option occur. From this angle of view it should be said that all stocks are having moments with high volatility and moments with small volatility.

Because the interest rate remains practically unchanged during the life time of an option contract, in order to see the stock price tunneling, the width of potential wall must be thin, and as result the volatility must be low. Logically, to have price tunnel out of the range bound the volatility must experience an important decrease in value, in a relative short period of time. Being computed constantly the fall in stock volatility should be easily observed in the market.

To conclude, **before a tunneling effect can occur in the stock market, a dramatic fall in the stock volatility, in a very short period of time, should be seen.** A surprising result that is completely different from the conventional market wisdom which states that in case implied volatility decreases the price of options usually decreases too.

To illustrate this intriguing, surprising but also very important result, two examples from the market practice were retained. Figure 5 captures tunneling effect of two well-known stocks price, namely Google (GOOG) and LinkedIn (LNKD), along with the predicted fall in the volatility.

Figure 5

Both pictures are explicitly showed that the moment of dramatically fall in the volatility coincides with an abrupt rise in the stock price.

6. **Transmission coefficient**

A more accurate result for tunneling probabilities latter discussed can be obtains by taken the distance d being fragmented in smaller pieces as in figure 6.

Figure 6

Summation over all this tiny regions gives the result:

$$T = e^{-2\sum_i (q_i d_i)} \qquad (21)$$

Taking the integral instead of the sum:

$$T = e^{-2\sqrt{\frac{r}{\sigma^4}(\sigma^2 + r)} \int_K^{S_1} \sqrt{\frac{1}{S^2} - \lambda} \, ds} \qquad (22)$$

Evaluating the integral in (30), from strike price K to the price outside the wall $S_1$ and rearranging the terms, the probability of price tunneling the potential wall is:

$$T = e^{-2\sqrt{\frac{r}{\sigma^4}(\sigma^2+r)} \left[\frac{1}{2}\ln\left(\left|\frac{\sqrt{1-\lambda K^2}+1}{\sqrt{1-\lambda K^2}-1}\right|\right) - \sqrt{1-\lambda K^2}\right]} \qquad (23)$$

Since $\lambda = \frac{r}{\sigma}$, the transmission coefficient is:

$$T = e^{-2\sqrt{\frac{r}{\sigma^4}(\sigma^2+r)}\left[\frac{1}{2}\ln\left(\left|\frac{\sqrt{1-\frac{r}{\sigma}K^2}+1}{\sqrt{1-\frac{r}{\sigma}K^2}-1}\right|\right)-\sqrt{1-\frac{r}{\sigma}K^2}\right]} \qquad (24)$$

and represent the probability of stock price to penetrate through the walls defined by support or resistance level, known as transmission coefficient.

The transmission coefficient formula may look complicated, but is easy to be compute since T depends only on known parameters: interest rate, volatility and strike price.

In the everyday trading practice on the market, quantum tunneling of stock price out of the range bound that it moved for some time, is a very common phenomenon. I guess there is no trader that shouldn't see, at least one time, a stock price effectively "exploding", rising very fast in a short period of time. Empirical evidence of stock tunneling could be find every day in the stock market, and not only. That makes this section purely informative. The interested reader can easily find many other examples on the financial market and compute transmission coefficient.

The Table 3 below gathers together some data for four different stocks, which were track from the site www.optionistics.com:

Table 3

The table is instructive by showing the way K, d and T are computed. Notice, also, the important fall in the volatility ahead of price tunneling.

## 7. Conclusions

Quantum tunneling effect is employ in explaining explosive violent stock price movements that sometime take place in range bound markets. The theoretical approach that extend quantum tunneling to financial markets it was made possible by a recently deduced time-independent equation in option valuation. The aforementioned equation is identical to the one-dimensional time-independent Schrodinger equation.

The shape of the potential in the financial time-independent equation and its form are an exact match of the nuclear α-particle decay as Gamow expressed it almost 100 years ago. Following this line of viewing the market price movements a consistent theory of financial quantum tunneling is build.

The probability of stock price tunneling out of the range bound, in the form of a financial adapted transmission coefficient is deduced. Transmission coefficient is depending on financial observables like stock volatility and interest rate and discussions around how these observables affect it can be found in the present paper.

Independent of transmission coefficient, another important phenomenon found in the present paper is that price tunneling effect is related to a dramatic fall in the stock volatility right prior a price penetration happened.

Market evidences of this important phenomenon are captured and shown in some suggestive figures.

**Tables**

| Interest rate -r | Volatility -$\sigma$ | Transmission coefficient- T(%) | Distance of penetration |
|---|---|---|---|
| 0.01 | 0.53 | 73 | 4.88 |
| 0.02 | 0.53 | 75 | 2.75 |
| 0.03 | 0.53 | 79 | 1.80 |
| 0.04 | 0.53 | 83 | 1.24 |
| 0.05 | 0.53 | 87 | 0.85 |
| 0.06 | 0.53 | 91 | 0.57 |
| 0.07 | 0.53 | 95 | 0.35 |

Table 1. Probability of tunneling and distance of penetration for different values of interest rate and constant volatility

| Interest rate -r | Volatility -$\sigma$ | Transmission coefficient- T(%) | Distance of penetration |
|---|---|---|---|
| 0.05 | 0.43 | 91 | 0.53 |
| 0.05 | 0.53 | 87 | 0.85 |
| 0.05 | 0.63 | 85 | 1.15 |
| 0.05 | 0.73 | 84 | 1.42 |
| 0.05 | 0.83 | 83.95 | 1.67 |
| 0.05 | 0.93 | 83.69 | 1.91 |
| 0.05 | 0.97 | 83.64 | 2.00 |

Table 2. Probability of tunneling and distance of penetration for different values of volatility and constant interest rate

| | date | r | $\sigma$ | Price at resistance | Price at support | K | d | T | Fall in volatility |
|---|---|---|---|---|---|---|---|---|---|
| LNKD - LINKEDIN CORP | 07-08.02.2013 | 0.03 | 0.47 | 127.2 | 123.3 | 3.9 | 0.058114 | 0.998675 | 0.63 to 0.39 |
| GOOG - GOOGLE INC A | 22-23.01.2013 | 0.03 | 0.15 | 704.7 | 702.6 | 2.1 | 0.136068 | 0.95 | 0.40 to 0.15 |
| HUM - HUMANA INC | 28.03-02.04.2013 | 0.03 | 0.31 | 70.08 | 66.95 | 3.13 | 0.08455 | 0.9948 | 0.43 to 0.25 |
| NFLX - NETFLIX INC | 23-24.01.2013 | 0.03 | 0.55 | 101.17 | 97.81 | 3.36 | 0.921744 | 0.933 | 0.95 to 0.55 |

Table 3. Transmission coefficient and fall in volatility prior price tunneling for some stocks.

**Figures**

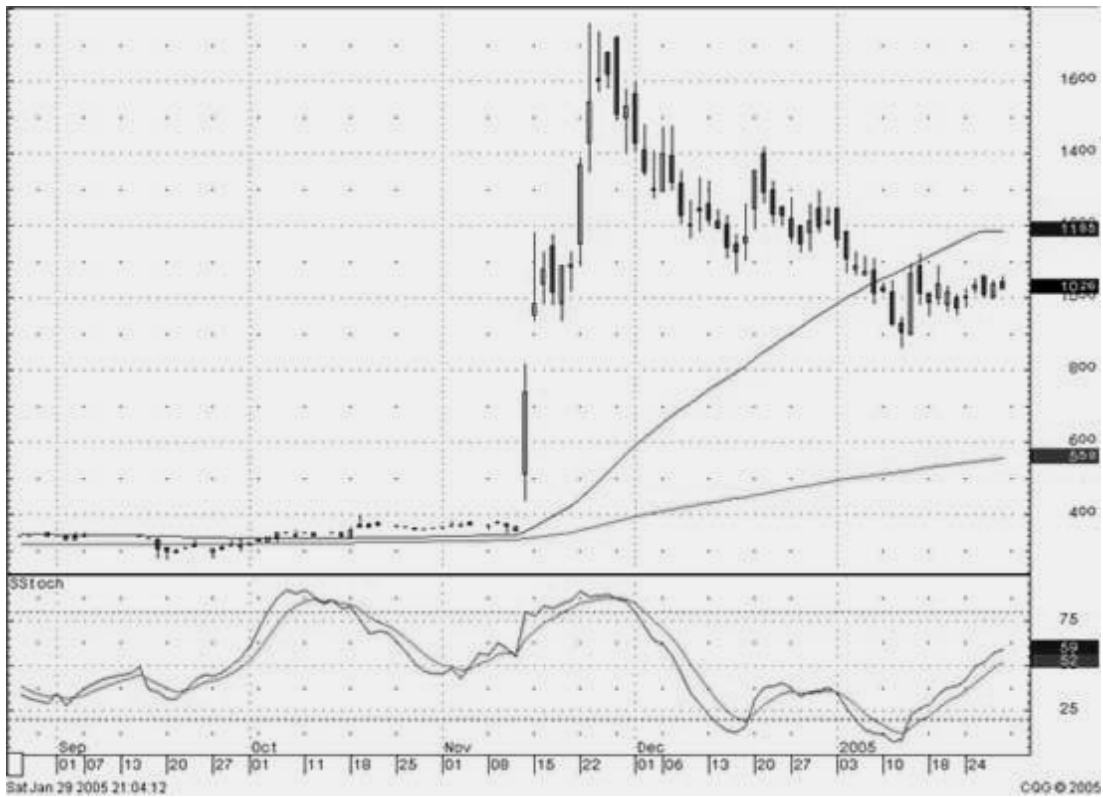

Figure 1. Picture taken from the trading platform illustrating the explosive violent move of the stock price out of the range bound.

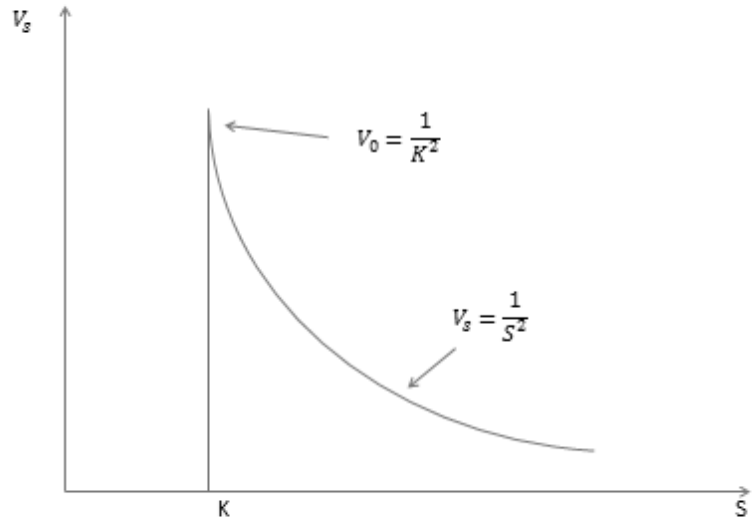

Figure 2. The shape of the stock price potential.

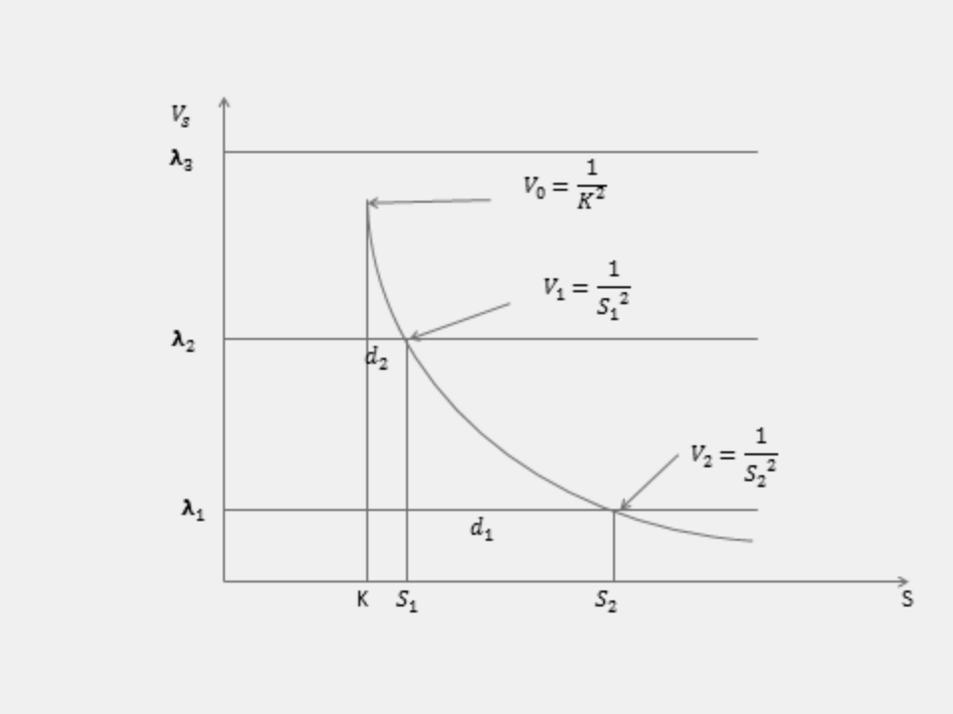

Figure 3. Different levels for **λ** constant.

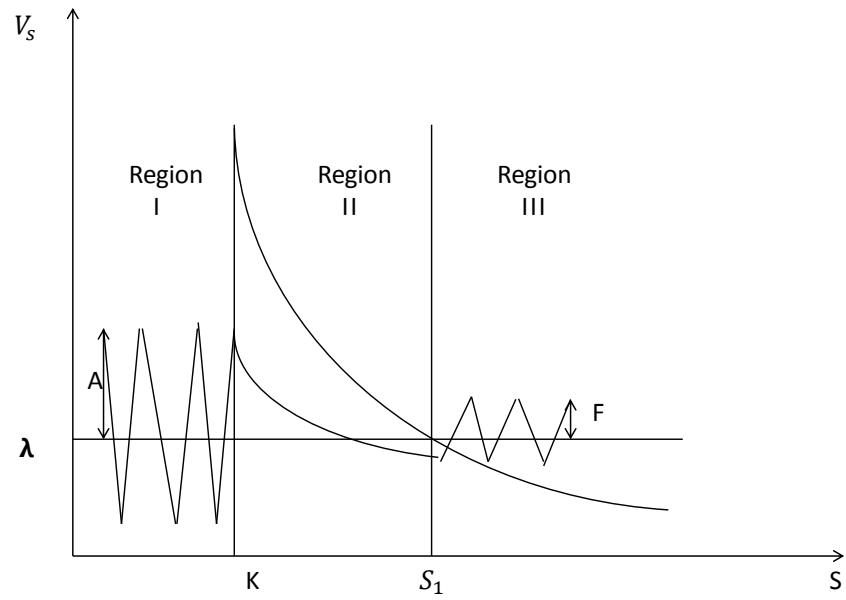

Figure 4. Behavior of $\psi_{(S)}$ function outside and inside the potential wall.

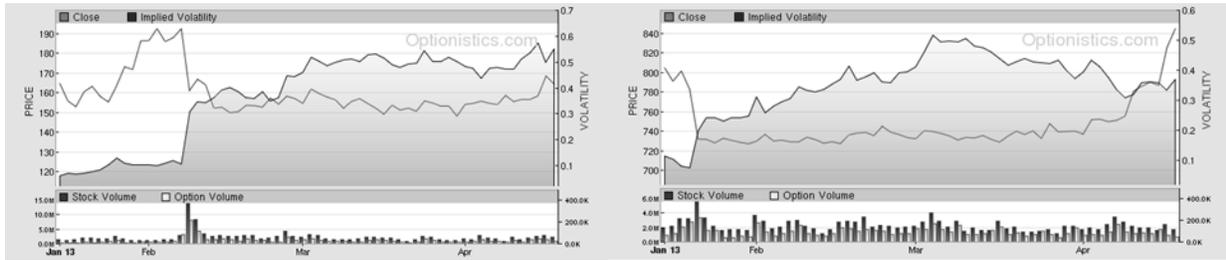

Figure 5. The fall in the stock volatility preceding the tunneling price effect for LNKD (left) and GOOG (right) stocks.

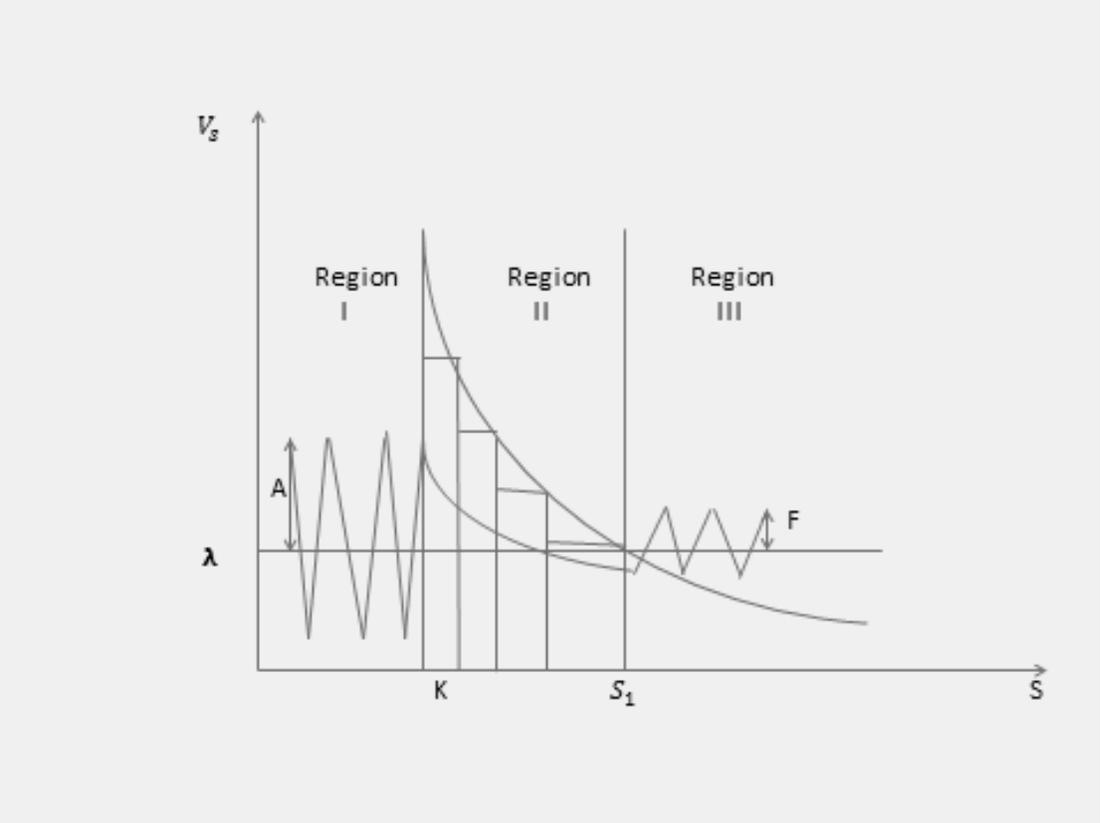

Figure 6. In region II inside the potential wall the penetration distance is fragmented in smaller regions.